# Ultra-stable lasers using hollow-core fibre

Zitong Feng[1,2,✉], Giuseppe Marra[2], Irene Barbeito Edreira[1,2], Hesham Sakr[1,3], Francesco Poletti[1], Radan Slavík[1]

[1]Optoelectronics Research Centre, University of Southampton, Southampton, SO17 1BJ, UK

[2]National Physical Laboratory, Hampton Road, Teddington TW11 0LW, UK

[3]Microsoft Azure Fiber, Romsey, SO51 9DL, UK

## Abstract

Ultra-stable lasers are fundamental to a growing range of applications, including optical frequency metrology, fundamental physics and quantum sensing. Their outstanding performance is achieved by stabilizing their frequency to Ultra-Low Expansion (ULE) optical cavities. However, the complexity of fabrication and assembly of these systems — even for compact designs — has been limiting their widespread deployment. While micro-resonators and optical fibre delay lines offer alternatives, their performance is significantly limited by thermally-induced frequency drift. Here we demonstrate, for the first time to the best of our knowledge, a laser stabilised to a Hollow Core Fibre (HCF) achieving comparable performance to ULE cavity-stabilised lasers. We achieve a frequency instability of $4.6\times10^{-15}$ at 1 s and a frequency drift of 88 mHz/s, reducible to 3.7 mHz/s with thermal correction. Furthermore, over 3-year characterization confirms the HCF's predictable long-term behaviour. These results and the simplicity of the HCF-based system pave the way to a high-performance and scalable solution for ultra-stable laser sources.

## Introduction

Ultra-stable lasers are at the heart of many of the world's most precise measurement devices, enabling a range of applications from optical atomic clocks[1,2], tests of fundamental physics[3,4], to precision spectroscopy[5,6], gravitational wave detectors[7-9], and gravity mapping[10]. The outstanding frequency stability of these laser systems is achieved by phase-locking an optical source to an external optical cavity. This technique has enabled remarkable fractional frequency instabilities reaching the $10^{-17}$ level, realized using either room-temperature, half-meter-scale (Figure 1, blue) ultra-low expansion (ULE) glass cavities[11,12], or shorter, cryogenic silicon cavities (Figure 1, purple)[13–16]. Achieving this level of frequency instability requires a tightly controlled environment, typically available in metrology laboratories.

Significant effort has focused on adapting optical cavity technology for field applications, such as portable optical atomic clocks[17-19], photonic-based radar systems[20], earthquake detection[21,22], and space missions for climate observation[23,24]. To meet this need, a variety of innovative ULE cavity geometries—including cubic[23,25-29], spherical[30,31], triangular[32], and cylindrical[24,27,33,34]—have been developed. By minimizing acceleration sensitivity and employing rigid, shock-resistant mounts, these 5–15 cm long cavities achieve frequency instabilities of $10^{-15}$ to $10^{-16}$ at 1 s (Figure 1, green). Furthermore, ULE cavities (≤ 1 cm) with improved manufacturability, when combined with fused silica mirror substrates, have demonstrated remarkable stability in the $10^{-14}$ to $10^{-15}$ range at 1s[35,36] (Figure 1, olive).

Other technologies, such as micro-resonators and optical fibre delay lines, have emerged as promising alternatives to optical cavities. They offer easier, more robust, and scalable construction at a lower cost, though they cannot yet match the frequency stability of ULE systems. Millimetre-scale micro-resonators are made using highly scalable fabrication and have achieved frequency instabilities of $10^{-13}$ at short time scales (10 ms)[37,38]. However, their high thermal sensitivity (e.g., 10 ppm/°C for silicon nitrate[37]) leads to large thermal drift in the kHz/s range (Figure 1, orange). This is 4–5 orders of magnitude larger than the tens of mHz/s drift typical for ULE cavities[27,28,33,34]. Lasers stabilized to optical delay lines made of a standard Single Mode Fibre (SMF) have achieved instabilities of $6.3\times10^{-15}$ at 16 ms[39]. However, similar to micro-resonators, the high thermal sensitivity of SMF (8 ppm/°C) degrades longer-term performance, with drifts of 28-270 Hz/s limiting the instability to $10^{-13}$ to $10^{-12}$ at 1 s[39]. Although placing the SMF in a vacuum chamber with multi-layer thermal shielding can lower the instability to $10^{-14}$ at 1 s[40,41], and post-processing to correct temperature fluctuations can even improve it to the $10^{-15}$ level (Figure 1, yellow)[41], this performance has only been demonstrated in measurements lasting a few hundred seconds. Over longer periods (5 days), the drift remains high at 16 Hz/s[42], about 2-3 orders of magnitude worse than ULE cavities. Additionally, unlike ULE cavities, SMF delay lines exhibit unpredictable frequency drift[42], potentially arising from relief of stress induced during fibre draw due to different properties of core and cladding glass materials.

In this work, we demonstrate for the first time that a laser stabilized to an HCF delay line can achieve frequency stability and drift comparable to that of medium-sized (5-15 cm) ULE cavity-stabilized lasers. These results are enabled by two key properties of HCF: its low thermal sensitivity (0.3 ppm/°C), as light propagates through a vacuum rather than a solid material, and its fabrication from a single glass material, which results in negligible built-in stress and far simpler drift behaviour than SMF. Over a 100-day measurement, we achieved a frequency instability of $4.6\times10^{-15}$ at 1 s and a frequency drift of 88 mHz/s (Figure 1, red point). This drift is comparable to ULE cavities and over two orders of magnitude lower than the best results from SMF systems. After removing the linear and thermally-induced trends, the residual frequency drift is less than 3.7 mHz/s. Furthermore, our characterization over a 3-year period reveals that the HCF's drift and thermal sensitivity decrease predictably and exponentially. This finding enables the prediction of required frequency drift corrections to maintain stability over many-year long time scales.

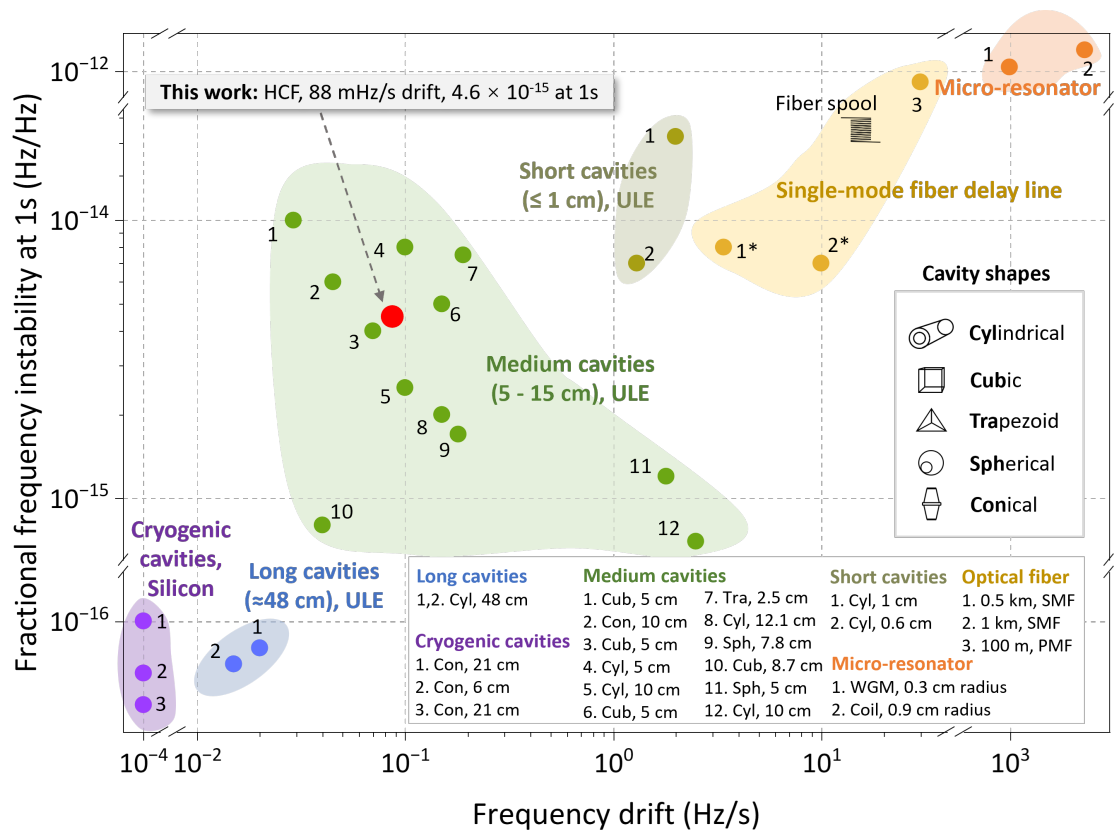

Figure 1. Comparison of frequency drift and frequency instability at 1s of laser stabilization to the state-of-the-art optical cavities, micro-resonators, and optical fibre delay lines. Purple area: silicon cavities at cryogenic temperatures[13-16]. Blue area: long length (around 0.5 m) ULE optical cavities[11,12]. Green area: Medium length (5 - 15 cm) ULE optical cavities[23-34]. Olive area: short length (≤ 1 cm) ULE optical cavities[35, 36]. Various shapes of cavity shown in the figure. Yellow area: single-mode optical fibre delay line[39-41]. Asterix refer to the performance after removing drift, which lasted only a few hundred seconds. Orange area: coil and whispering gallery mode (WGM) micro-resonators[37,38]. Red point: here presented our work by stabilizing lasers to an HCF delay line. The lists in the figure also display detailed information about various optical cavities, micro-resonators, and optical fibre delay lines, such as their dimensions.

## Experimental setup

Figure 2 shows the experimental setup consisting of a Michelson interferometer (MI) with a 300 m arm imbalance, achieved by inserting a 150 m-long HCF spool (the delay line) into one of the two arms. The HCF used is the Nested Antiresonant Nodeless Fiber (NANF)[43] developed at the Optoelectronics Research Centre, University of Southampton. The HCF was wound on a collapsible drum with 12 cm diameter, allowing the resulting fibre coil to remain unsupported. This design minimises stress-induced changes in optical path length that arise from the interface between the fibre and a supporting structure. The light from the laser source used to interrogate the delay line interferometer was coupled into the HCF using fibre-pigtailed collimators and focusing lenses. The coupling losses between SMF and HCF were measured to be 1 dB per connection, which could be reduced with an all-fibre solution (Discussion). At the output of both arms of the interferometer,

Faraday mirrors were installed to ensure polarization insensitivity of the interferometric fringe contrast. The delay line interferometer was placed in a vacuum environment to isolate it from external temperature and pressure changes. More details on the experimental setup are provided in the Methods.

To characterize the frequency stability of the HCF delay line optical reference, we employed two separate measurement methods (shown in Figure 2) that provide better measurement stability over long (Method 1) and short (Method 2) time scales, respectively. In Method1, we phase locked a laser (RIO, Luna Inc.) to an Optical Frequency Comb (OFC), then used it to interrogate the HCF delay line interferometer, converting the detected interferometric fringes into laser frequency fluctuation (Method section) to calculate the resulting frequency instability of a laser locked to the HCF delay line. Although Method1 provides sufficient long-term stability (Methods section), its measurement is limited at the $10^{-13}$ level for timescales shorter than 100 s, due to the stability of the RF source used to lock the repetition rate of the OFC. To address this limitation, Method2 used a ULE-cavity stabilized laser developed at NPL with short-term stability about two orders of magnitude better than the OFC used in Method1 (Method section). We phase-locked a laser source to the HCF optical delay line interferometer using the Pound-Drever-Hall technique and combined it with the ULE-cavity stabilized laser, measuring changes in their beat frequency (centred at approximately 30 MHz) after photodetection using a dead-time free frequency counter with a gate time of 1 s.

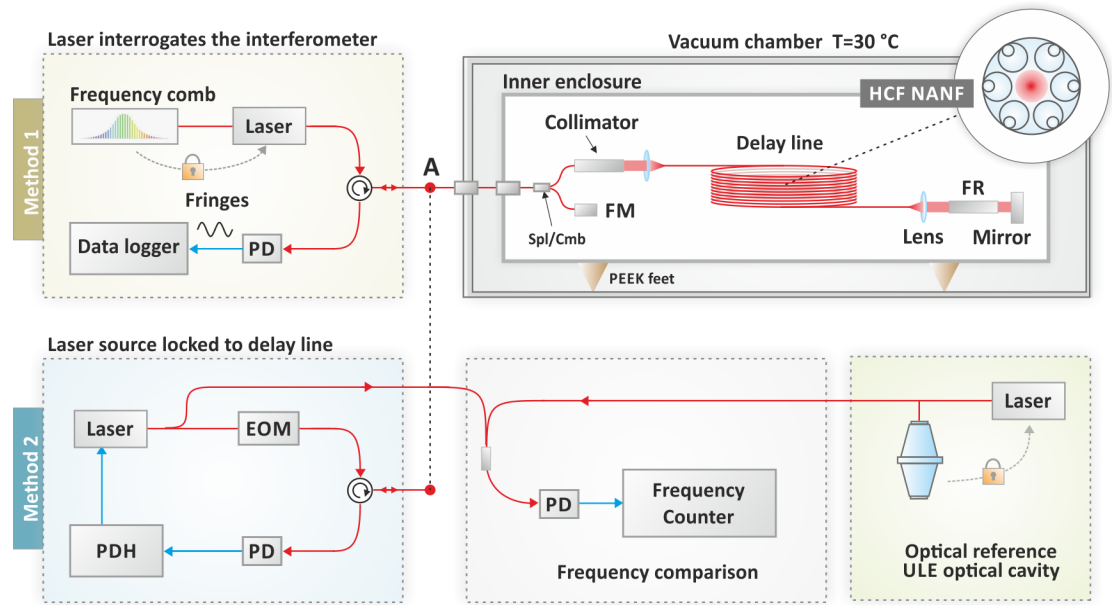

Figure 2. Experimental set-up. Vacuum chamber: HCF delay line arrangement within the vacuum environment; Method 1: Frequency stability of the HCF delay line measurement by recording the output intensity of the HCF delay line interferometer; Method 2: Locking a laser to the HCF delay line interferometer and comparing it with a laser stabilized by a ULE-Cavity. Point A: Input/Output of light to HCF delay line, FR: Faraday Rotator, FRM: Faraday Mirror, OC: Optical Coupler, EOM: electro-optical modulator-provide phase modulation for PDH locking.

## Laser frequency stability

Figure 3a shows the measured long-term frequency stability obtained using Method1. These measurements ran nearly uninterrupted over 109 days, with only two short gaps of a few hours (marked on Figure 3a) due to occasional OFC unlock. The measured frequency shows a near-monotonic increase over time, leading to a total frequency change of 836 kHz over the measurement duration. A linear fit of the frequency increase over this time frame indicates a drift of approximately 88 mHz/s. We attribute this frequency drift to aging of the silica material and its acrylate coating applied during the HCF fabrication to protect the fibre. We believe this is the first experimental observation of optical fibre aging, which appears to be of comparable level to ULE-cavities.

By removing the linear component of the frequency changes, we observe that the residual frequency changes closely match the behaviour of the temperature measured by a thermistor placed close to the HCF. The maximum residual frequency change observed is approximately 60 kHz (Figure 3b). By applying a bivariate regression algorithm, designed to minimize the ratio between the measured frequency fluctuations and temperature reading, we determined the thermal sensitivity of the HCF to be 0.32 ppm/K. This value, which closely matches previously reported results[44], was then used to subtract frequency changes due to temperature fluctuations (temperature corrected), as shown in Figure 3c. This reduced the peak-to-peak variation of the residual frequency fluctuations by a factor of 30, down to 2 kHz, and reduced the frequency drift to below 3.7 mHz/s.

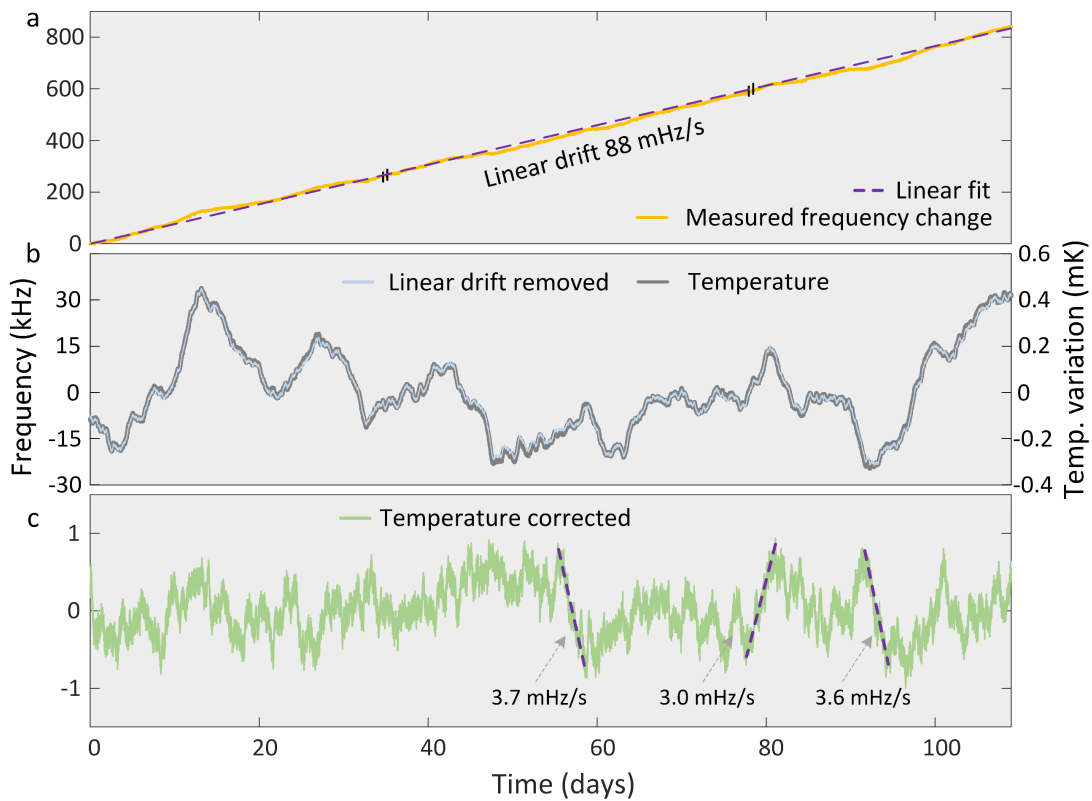

Figure 3. Frequency variation over 109 days period. a) Frequency variation with its linear fit, also showing two brief time periods (1.6 and 2.3 hours long) when the OFC signal was not available; b) Frequency variation after removing linear drift and its comparison with temperature measurement, showing their strong correlation; c) Residual frequency variation after removing variations due to temperature fluctuations, showing worst-case (fastest change) frequency drift at three time periods (>3 days).

Figure 4 shows Allan deviations of the fractional frequency instability for (i) the measured frequency changes shown in Figure 3a, (ii) for the same data with the linear component of the frequency drift removed, as in Figure 3b, and (iii) with further temperature-induced frequency changes removed, as in Figure 3c, all marked by open and half-solid circles. Figure 4 also shows results obtained with Method2 for time scales below 4000 s (solid and half-solid circle). There is a good agreement between both methods (half-solid circle) from 100 s to 4000 s averaging time. Below 100 s, Method1 is degraded by insufficient short-term stability of the OFC (red dash line), while the stability measured with Method2 reached $4.6\times10^{-15}$ at 1 s.

The results show the stability is limited by the frequency drift for averaging times longer than 10 s, reaching $3.9\times10^{-11}$ at 1 day and $4.3\times10^{-9}$ at 109 days (3.7 months). After removing the linear component of the frequency drift and temperature-induced frequency changes, the instability for these time scales improves by approximately two orders of magnitude.

The grey shaded area shows the performance range of a number of ULE cavity-stabilized systems, demonstrating that the HCF-based stabilization system achieves a comparable level of frequency stability.

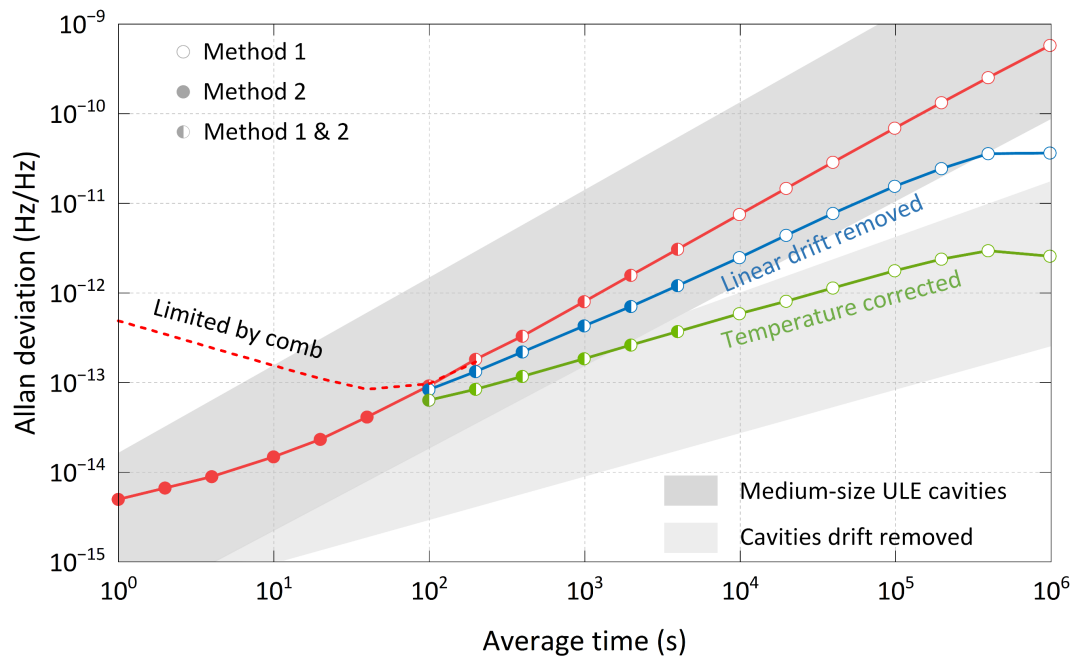

Figure 4. Allan deviation of fractional frequency fluctuations over $10^6$ s time scales for the HCF stabilised laser, measured using Method1 and Method2. Open circle: measured by Method1 only; Solid circle: measured by Method2 only; Half-solid circle: both Methods used. Dash red line: measured by Method1 to show the limitation of OFC; The three trace colours correspond to the data as total frequency variations (red), with linear drift removed (blue), and with both linear drift and temperature effects removed (green). The two grey regions indicate the typical performance ranges for ultra stable lasers based on medium-sized ULE cavities for both drifting (darker grey) and drift-compensated systems (lighter grey).

To investigate the source of instability at short time scales, we measure the phase noise at offset frequencies between 1 Hz and 10 kHz by reducing the counter gate time to 50 μs. The power spectral density of the measured phase noise, shown in Figure 5, reveals an overall trend following a $1/f^2$ power-law, corresponding to a random walk in laser phase. An integrated linewidth of 0.6 Hz is obtained by integrating the phase noise over the 1 Hz to 10 kHz. From 10 to 100 Hz, we observe a departure from the $1/f^2$ slope with excess noise and discrete components. By comparing the measured phase noise with environmental acceleration noise recorded by a nearby seismometer, we confirm that the observed discrete tones and the bump originate from the acceleration-induced noise of the HCF delay-line interferometer. To estimate the expected frequency stability without excess noise due to acceleration, we assume that the phase noise follows perfectly the $1/f^2$ random walk of phase power-law, as shown by the purple dash line in Figure 5. From the -8 dBrad$^2$/Hz at 1 Hz we can calculate an instability of $2.2\times10^{-15}$ at 1 s[45]. This improved level of stability could potentially be achieved in future tests by installing the experimental setup on a passive/active vibration isolation platform.

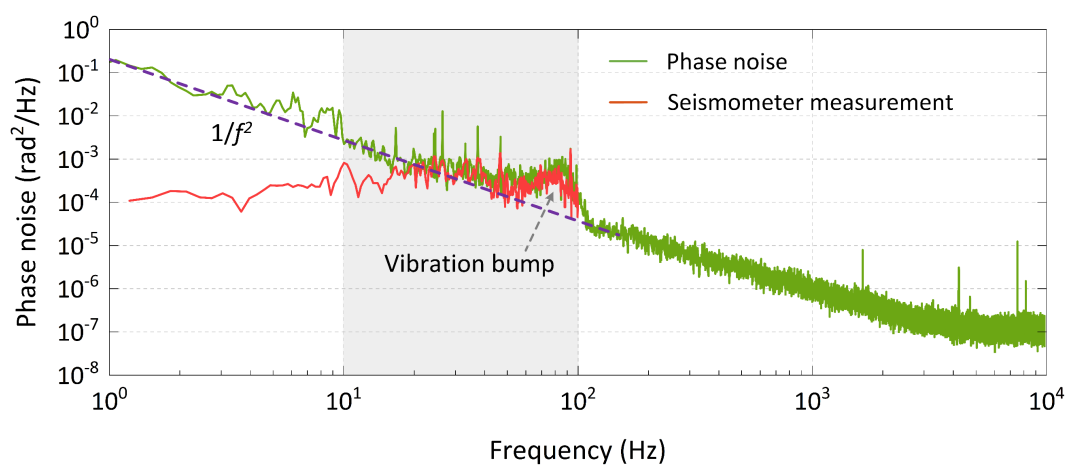

Figure 5. Phase noise (green) and seismometer measurement (red). The bump and discrete noise components in the grey-shaded region are caused by vibrations.

## Long-term behaviour of thermal sensitivity and frequency drift

We have shown that the thermally-induced changes in the HCF delay line optical length can be efficiently compensated by precise temperature readings with a thermistor installed close to the HCF spool. However, this compensation can only be successfully performed if the HCF thermal sensitivity is known with sufficient accuracy. We operated our experimental setup continuously over the course of 3 years, allowing us to characterize how the HCF thermal sensitivity changes over time, Figure 6a. During this long observation time, we measured an exponential decrease of the thermal sensitivity. By fitting an inverse exponential function, we calculated a time constant of 401 ± 24 days. According to this fit (Figure 6a, green dashed line), the thermal sensitivity is projected to asymptotically approach 0.3 ppm/K. This value is consistent with the reported coefficient of thermal expansion for bare (uncoated) HCF[46], but lower than expected for a coated HCF (e.g., 0.52 ppm/K reported in ref. 44). To gain a better understanding of our observations, further investigations are needed, e.g., a comparison of standard-coated, thinly-coated, and bare

HCFs over an extended time period in vacuum would indicate the effect of coating thickness on thermal sensitivity changes driven by air and moisture egress.

We also characterized the HCF frequency drift over three years, Figure 6b. Similarly to thermal sensitivity, the drift change fits well with an exponential function (red dash line) with a time constant of 455 ± 20 days. Based on this exponential fit, we predict that the drift will improve to 70 mHz/s after a further 1 year and lower than 50 mHz/s after a further 4.5 years.

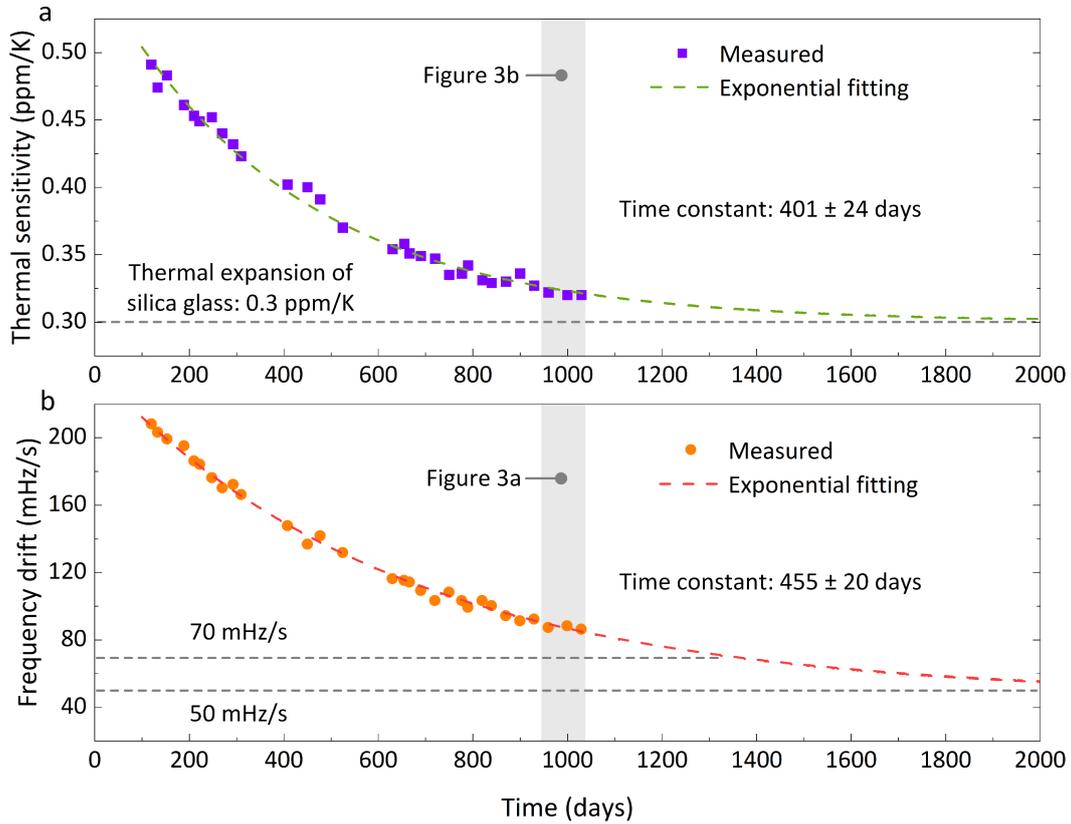

Figure 6. (a) thermal sensitivity and (b) frequency drift changes of the HCF delay line measured over a period of 3 years. Gray area corresponds to frequency drift shown in Figure 3a, and thermal sensitivity used in Figure 3b to remove temperature effect.

The predictable behaviour of both the thermal expansion and frequency drift indicate that the compensation of the frequency changes caused by these effects can be performed accurately over very long timescales. Hence, we expect that the performance of the HCF-based stabilization system could be maintained over many years of operation.

## Discussion

We demonstrated an ultra-stable laser system based on an HCF delay line, with frequency stability performance achievable today only with bulk-optics-based cavity-stabilised systems. Our HCF-stabilized laser delivered a fractional frequency instability of $4.6\times10^{-15}$ at 1s and an 88 mHz/s frequency drift—2 orders of magnitude better than other optical fibre system and 3 orders better than reported micro-resonators. This performance was maintained for over 100 days during our tests.

We further characterized the thermal expansion and the frequency drift of the HCF over 3 years and observed that both effects exhibit an exponentially decreasing behaviour. This allows prediction of the HCF frequency drift, enabling effective drift compensation over several years. In our work, a residual frequency drift below 3.7 mHz/s was obtained after drift compensation. Phase noise measurement indicates that short-term frequency instability was primarily limited by environment-induced vibrations, with integration yielding a linewidth of 0.6 Hz. In future work, this excess noise could be reduced by adding an active vibration isolation platform, designing a vibrationally-insensitive spool[40], or measuring and real-time cancellation of vibration-induced phase noise using a seismometer[47]. Possible limitations may also arise from Residual Amplitude Modulation in the phase modulator used in PDH locking, which, at a typical level of 100 ppm for Ti-diffused waveguides in $LiNbO_3$, can limit frequency instability to $1.6\times10^{-15}$ at 1 s. However, this effect could be suppressed below $10^{-15}$ by implementing active servo feedback[48].

We estimate that the HCF spool diameter could be reduced to 1-2 cm with negligible increase in loss[49], enabling significant system compactness. Future implementations may also benefit from connectorized HCF, eliminating the need for alignment into the fibre. Advances in connection of HCF with SMF have shown a loss below 0.1 dB and reflection levels below -60 dB[50]. This all-fibre solution, which removes the reliance on free-space optics, enhances the system's robustness, reducing the risk of misalignment or structural failure caused by environmental perturbations and mechanical vibrations. This is particularly beneficial for space missions, where maintaining performance under launch conditions with accelerations more than 10 g is critical.

Recent developments of the HCF technology have demonstrated lower attenuation in the 1550 nm region (<0.11 dB/km[51]) than the best SMF (0.14 dB/km[52]) available today, making HCF the lowest attenuation optical fibre ever realised. These achievements are speeding up the commercialization of HCF technology, leading to higher production volumes, which in turn are expected to drive a significant increase in availability and reduction in cost. In this work, the stabilization technique was demonstrated at a 1543 nm wavelength. However, it could easily be extended to other wavelength regions. HCF can be designed for a very wide range of wavelengths, from the edge of mid-IR (~2000 nm) to the visible range[51, 54, 55].

In contrast to ULE cavity-based laser stabilization systems, the HCF-based solution we have demonstrated does not require precise machining of specialized glass materials, the manufacture of high reflectivity mirrors, their precise optical contacting, and the skilled cavity mode-finding alignment procedure. This allows a higher scalability of the solution and, thus, a wider range of applications. Whilst optical cavities, especially those of larger size, are likely to continue to be the solution of choice for the most demanding frequency stability performance for years to come, the simplicity and ease of construction of the HCF-based system presented in this work paves the way to a faster adoption of ultra-stable laser sources across a wider range of applications than has been possible so far.

# Methods

**HCF sample used**

6-tube HCF of Nested Antiresonant Nodeless Fibre (NANF) geometry manufactured in 2019. It has a 37 μm core diameter, single acrylate coating of 338 μm outer diameter, and loss of 2.5 dB/km at 1550 nm.

**MI interferometer set-up in vacuum chamber**

Figure 2 shows the MI interferometer built in a vacuum chamber. The laser light input at A point is split into two paths by a 20/80 SMF-based fused coupler. The 80% coupler output is directed into the HCF via collimating and focusing lenses. After transmission through the HCF, the output light is collimated and directed onto a Faraday rotator, then reflected back by a mirror. The 20% coupler output serves as a reference, returning directly through a fibre-pigtailed Faraday rotator mirror. The light from both branches is combined to generate the interferometric fringes. High interference contrast is achieved with the output power from both interferometer branches approximately equal, as HCF branch loss is offset by launching higher power via the 80% port of the 20/80 coupler. The Faraday-mirror-assisted MI configuration effectively removes contributions of polarization changes to the laser frequency stability as well as effects of polarization dependent loss and polarization mode dispersion.

The MI was housed in an aluminium box, supported by four PEEK (thermal conductivity 0.26 W m$^{-1}$ K$^{-1}$) rods with conical bases to minimize contact with the vacuum chamber for thermal isolation. The vacuum chamber was first evacuated using a turbo vacuum pump and subsequently maintained at the 10$^{-5}$ mbar level throughout the measurement duration by an ion pump. The vacuum chamber was temperature stabilized to 30 °C and encased in a Celotex-insulated box. The gap between this insulating enclosure and the chamber was filled with refrigerant gel-packs, extending the thermal time constant to approximately two days. The temperature was stabilized to within 1 mK around the HCF spool over several months.

**Optical frequency comb (OFC)**

The OFC, synchronized to an RF reference with an ultra-low phase noise oscillator and further referenced to a GPS-based atomic clock, facilitated a stability transfer that enabled measurements of frequency at fractional instabilities of 10$^{-13}$ at 1 s and 10$^{-12}$ at several days. This performance is nearly an order of magnitude better than that of HCF delay line for long averaging times (days), sufficient to measure frequency drift of the HCF-stabilized laser.

**ULE-Cavity stabilized laser**

The cavity-stabilized laser used for short-term frequency characterisation was a transportable ultra-stable laser system developed by the National Physical Laboratory (NPL) and is based on a 10 cm ULE cavity with an instability of 10$^{-15}$ at 1 s and a linear drift of 40 mHz/s. The details have been published in [33].

**Convert MI fringe to frequency**

The fringe voltage signal $V(t)$, measured from the photodetector, was first corrected for any offset $V_{offset}$ and normalized by the maximum voltage $V_{max}$. This normalized signal is directly related to the phase difference $\Delta\varphi(t)$, which reflects changes in the optical path length between the interferometer arms. $\Delta\varphi(t)$ is unwrapped to ensure continuity. The frequency fluctuation $\Delta f(t)$ is then derived by differentiating the unwrapped phase with respect to time, with the result normalized by $\frac{1}{2\pi}$ to obtain the frequency change:

$$\Delta f(t) = C_{calib} \cdot \frac{1}{2\pi} \frac{d}{dt}\left[Unwrap\left(2arccos\left(\frac{V(t) - V_{offset}}{V_{max} - V_{offset}}\right)\right)\right] \quad (1)$$

To ensure accurate voltage-to-frequency conversion, a calibration factor $C_{calib}$ was determined by comparing the measured fringe voltage to known frequency changes from an OFC locked laser. The calibration corrects for any scaling differences between the voltage signal and the actual frequency changes:

$$\Delta f_{calibrated}(t) = C_{calib} \cdot \frac{1}{2\pi} \cdot \Delta V_{fringe} \quad (2)$$

# Acknowledgements

This work was financially supported by Engineering and Physical Sciences Research Council (EP/P030181/1); VACUUM (EP/W037440/1) and by the UK government Department for Science Innovation and Technology through the National Measurement System Programme.

**Conflict of interest**

The authors declare that they have no conflict of interest.

**Data availability**

All data are available in the main text or in the "Methods" section.